\newif\ifdraft

\ifdraft
\documentclass[preprint,12pt,authoryear,fleqn]{elsarticle}
\else
\documentclass[final,3p,twocolumn,times,authoryear,lefttitle,fleqn]{elsarticle}
\fi

\usepackage{alphabeta}
\usepackage{amsmath,amssymb}
\usepackage[hyphens]{url}
\usepackage[hidelinks]{hyperref}
\usepackage[separate-uncertainty=true,multi-part-units=single]{siunitx}
\usepackage{lineno}
\usepackage{multirow}
\usepackage{floatrow}
\usepackage{orcidlink}

\journal{Applied Radiation and Isotopes}

\begin{document}

\begin{frontmatter}

\title{Conceptualization of electronic dose to water and water-equivalent gas for ionometric dosimetry in external beam radiotherapy}

\author{Nobuyuki Kanematsu\,\orcidlink{0000-0002-2534-9933}}
\ead{kanematsu.nobuyuki@qst.go.jp}
\address{Department of Accelerator and Medical Physics, Institute for Quantum Medial Science, National Institutes for Quantum Science and Technology (QST), 4-9-1 Anagawa, Inage-ku, Chiba 263-8555, Japan}

\begin{abstract}
\small
Electronic dose to water, which is the radiant energy imparted to a unit mass of water by electronic collisions, is conceptualized to exclude corrections and uncertainties associated with energy transfer to atomic nuclei and to enhance medical relevance for radiotherapy dosimetry. 
Based on the international code of practice, the procedures to determine electronic doses were formulated for high-energy photon, electron, proton, and ion beams.
Nitrogen-based water-equivalent gas (WEG) mixtures were designed for use in gas-sealed ionization chambers for proton and ion beams.
A thought experiment of reference dosimetry for electronic dose demonstrated its compatibility with absorbed dose for photon and electron beams.
For proton and ion beams, the transition from absorbed dose to electronic dose will change apparent dose values by \qty{-1.4}{\%} and \qty{-2.2}{\%}, and decrease the uncertainty from \qty{1.6}{\%} to \qty{1.5}{\%} and from \qty{2.6}{\%} to \qty{2.1}{\%}, respectively.
With WEG ionization chambers, the uncertainty will further decrease to \qty{1.2}{\%} for proton beams and \qty{1.5}{\%} for ion beams.
The formulated framework will be applicable to radiotherapy dosimetry with traceable dosimeter calibration under the international radiation dosimetry system.
\end{abstract}

\begin{keyword}
\small
radiation dosimetry \sep ionization chamber \sep water-equivalent gas \sep proton beam therapy \sep ion beam therapy
\end{keyword}

\end{frontmatter}

\section{Introduction}

Absorbed dose to water has long been used as a standard quantity for medical radiation doses.
In general, the doses of high-energy radiotherapy beams are measured electrically using ionization chambers.
Under careful influence control or correction for temperature, pressure, ion recombination, etc., the ionization charge $M$ is related to the dose $D$ by
\begin{equation}
D = \frac{M}{e}\,\frac{{W_\mathrm{air}}_Q}{\rho_\mathrm{air}\,V_\mathrm{ch}}\,{s_\mathrm{w/air}}_Q\,{p_\mathrm{ch}}_Q,
\label{eq_D}
\end{equation}
where $M$ is divided by the elementary charge $e$ to count the number of ion pairs formed, multiplied by the mean energy expended in air per ion pair formed ${W_\mathrm{air}}_Q$ to measure the energy imparted to air in the chamber, divided by the air density $\rho_\mathrm{air}$ and the chamber volume $V_\mathrm{ch}$ to derive the absorbed dose to air, multiplied by the water-to-air mass stopping power ratio ${s_\mathrm{w/air}}_Q = {s_\mathrm{w}}_Q/{s_\mathrm{air}}_Q$\footnote{Notation: $x_{a/b} = x_a / x_b$ is the ratio of two subscripted instances of quantity $x$.} to convert to the absorbed dose to water, and multiplied by the overall chamber perturbation factor ${p_\mathrm{ch}}_Q$ to correct all dosimetric effects of the chamber structure.
The beam quality $Q$ represents the spectrum of beam particles. 

The International Atomic Energy Agency (IAEA) established and recently updated the international code of practice (ICP) as the standard protocols for dosimetry in external beam radiotherapy \citep{IAEA_2024}.
Since the individual chamber volume $V_\mathrm{ch}$ is not precisely known, the proportional relation $D \propto M$ in Eq.~(\ref{eq_D}) is calibrated to the absorbed dose to water given by a reference absolute dosimeter in a reference beam of quality $Q_0$, which is $^{60}$Co \gamma\ rays.
The dosimeter calibration coefficient ${N_\mathrm{D}}_{Q_0} = (D/M)_{Q_0}$ is determined for individual ionization chambers at secondary standards calibration laboratories. 
In dosimetry for an arbitrary beam of quality $Q$, the dose is given by
\begin{equation}
D = M\,{N_\mathrm{D}}_{Q_0}\,k_{Q/Q_0}, 
\label{eq_Dref}
\end{equation}
where the beam-quality correction factor $k_{Q/Q_0}$ is a product of three cause-specific factors,
\begin{equation}
k_{Q/Q_0} = {W_\mathrm{air}}_{Q/Q_0}\,{s_\mathrm{w/air}}_{Q/Q_0}\,{p_\mathrm{ch}}_{Q/Q_0}.
\label{eq_kQ}
\end{equation}
The $k_{Q/Q_0}$ factors for standard dosimeter models are listed for practical use in data tables of ICP \citep{IAEA_2024}.
They were evaluated for the spectrum of particles per beam quality $Q$ by Monte Carlo simulation using standard physical data.

The International Commission on Radiation Units and Measurements (ICRU) assigned single values per beam type for the standard $W_\mathrm{air}$ values, despite possible energy dependence \citep{ICRU_2016}.
They are \qty{33.97 \pm 0.12}{eV} for electrons in electron and photon beams including $^{60}$Co \gamma\ rays, $\qty{34.44 \pm 0.14}{eV}$ for proton beams, and $\qty{34.71 \pm 0.52}{eV}$ for ion beams covering helium to neon. 
The common $W_\mathrm{air}$ value for photon and electron beams leads to zero uncertainty for the ratio ${W_\mathrm{air}}_{Q/Q_0} = 1$. 
The primary particles of proton and ion beams coherently slow down in matter, resulting in significant variation in ${s_\mathrm{w/air}}_{Q/Q_0}$ with large uncertainty for the correction under limitations in the modeling of beam quality.

Table~\ref{tab_uncertainties} shows the relative standard uncertainties of $k_{Q/Q_0}$, its cause-specific constituents, and ionization measurement $M$ with cylindrical chambers in reference dosimetry for single-energy fields \citep{IAEA_2024}, which lead to the relative standard uncertainty for dosimetry,
\begin{equation}
\hat\Delta D = \frac{\Delta D}{D} = \sqrt{(\hat\Delta M)^2 + (\hat\Delta {N_\mathrm{D}}_{Q_0})^2 + (\hat\Delta k_{Q/Q_0})^2},
\label{eq_DeltaD}
\end{equation}
where $\hat\Delta {N_\mathrm{D}}_{Q_0} = \qty{0.6}{\%}$ has been estimated as a typical uncertainty of dosimeter calibration coefficient \citep{IAEA_2024}.\footnote{Notation: $\hat\Delta x = \Delta x/|x|$ is the relative ($\hat\ $) standard uncertainty ($\Delta $) of quantity $x$.}
The dose uncertainty $\hat\Delta D$ for proton and ion beams (\qty{1.6}{\%} and \qty{2.6}{\%}) is significantly larger than that for photon and electron beams (\qty{1.0}{\%} and \qty{1.1}{\%}).

\begin{table}[htb]
\caption{Relative standard uncertainties of beam-quality correction factor $k_{Q/Q_0}$ and its cause-specific constituents (mean energy expended in air per ion pair formed $W_\mathrm{air}$, water-to-air mass stopping power ratio $s_\mathrm{w/air}$, and overall chamber perturbation factor $p_\mathrm{ch}$ ), and typical uncertainties of ionization measurement $M$ in reference dosimetry with cylindrical chambers for single-energy fields \protect\citep{IAEA_2024}.}
\label{tab_uncertainties}
\begin{tabular}{lccccc}
\hline\hline
\multirow{2}{*}{Beam type} & \multicolumn{5}{c}{Relative standard uncertainty ($\hat\Delta$)} \\
\cline{2-6}
& $k_{Q/Q_0}$ & ${W_\mathrm{air}}_{Q/Q_0}$ & ${s_\mathrm{w/air}}_{Q/Q_0}$ & ${p_\mathrm{ch}}_{Q/Q_0}$ & $M$ \\
\hline
Photon & \qty{0.62}{\%} & \qty{0}{\%} & \multicolumn{2}{c}{\qty{0.62}{\%}\mpfootnotemark[1]} & \qty{0.56}{\%} \\
Electron & \qty{0.68}{\%} & \qty{0}{\%} & \multicolumn{2}{c}{\qty{0.68}{\%}\mpfootnotemark[1]} & \qty{0.56}{\%} \\
Proton & \qty{1.4}{\%} & \qty{0.53}{\%} & \qty{1.08}{\%}\mpfootnotemark[2]\mpfootnotemark[3] & \qty{0.66}{\%}\mpfootnotemark[2] & \qty{0.60}{\%} \\
Ion & \qty{2.4}{\%} & \qty{1.54}{\%} & \qty{1.56}{\%}\mpfootnotemark[2]\mpfootnotemark[3] & \qty{1.04}{\%}\mpfootnotemark[2] & \qty{0.76}{\%} \\
\hline\hline
\end{tabular}
\footnotetext[1]{Fully evaluated in the combined form $\hat\Delta (s_\mathrm{w/air}\ p_\mathrm{ch})_{Q/Q_0}$.}
\footnotetext[2]{Including a quadratic half share of $\hat\Delta (s_\mathrm{w/air}\,p_\mathrm{ch})_{Q_0} = \qty{0.4}{\%}$.}
\footnotetext[3]{Including additional \qty{0.3}{\%} to $\hat\Delta {s_\mathrm{w/air}}_Q$ for beam-quality assignment.}
\end{table}

This study aims to minimize the dosimetric uncertainty for proton and ion beams by a new dosimetry framework, which will consistently cover photon and electron beams as well.
In the following sections, we review the standard radiation dosimetry to identify its limitations, conceptualize a new dose quantity as a solution, and propose new ionization chamber gases to improve the dosimetric accuracy for proton and ion beams. 
Based on ICP as the gold standard, we formulate the dosimetry procedures for high-energy photon, electron, proton, and ion beams. 
We evaluate their feasibility and significance by a thought experiment of reference dosimetry.

\section{Materials and methods}

\subsection{Conventional dosimetry framework} \label{sec_conventional}

\subsubsection{Radiotherapy dosimetry}
The kinetic energy of charged particles in a radiation beam is imparted to matter by collisions with atomic electrons in electronic stopping for atomic excitation and ionization, and with atomic nuclei in nuclear stopping for nuclear recoil \citep{ICRU_2016}.
The electronic and nuclear stopping powers, $S_\mathrm{e}$ and $S_\mathrm{N}$, constitute the stopping power $S$ for dosimetry, while the radiative stopping power is excluded because secondary radiation does not directly contribute to dose but is incorporated into the beam to modify the beam quality.

The energy absorbed into matter will eventually convert to molecular kinetic energy or heat after ion recombination, atomic de-excitation, and photon absorption processes, except for the energy expended for material modification or heat defect. 
In calorimetric absolute dosimetry, the heat defect for chemical reactions are precisely controlled or corrected \citep{Klassen_1997}.
However, the heat defect for endothermic nuclear reactions, such as nuclear transmutation $^{16}$O\,(p,\,p\,n)$^{15}$O with the reaction energy $Q = \qty{-15.7}{MeV}$, is ignored despite its magnitude and prevalence in proton and ion beams \citep{Parodi_2018}.
In fact, absorbed dose has been designed to exclude such nuclear heat defect \citep{ICRU_2016}, which warrants calorimetric dosimetry without correction.

Clinical dosimetry typically employs standard air ionization chambers, whose ionization charge has been calibrated to reference absorbed dose traceable to the international standards \citep{IAEA_2024}.
In the Bragg--Gray cavity theory, it is assumed that all charged particles penetrate a small cavity to deposit some energy locally and that no other interactions occur in the cavity.
Then, the dose to air is proportional to the mass stopping power of air, and the water-to-air mass stopping power ratio $s_\mathrm{w/air} = (S/\rho)_\mathrm{w/air}$ is a correction factor to derive absorbed dose to water in ionometric dosimetry.
The nuclear stopping power $S_\mathrm{N}$ is insignificant for electrons because they are too light to transfer their energy to heavy nuclei.
In the ratio $s_\mathrm{w/air}$ with partial cancellation between water and air, the contribution of $S_\mathrm{N}$ has also been disregarded for protons and ions \citep{Medin_1997,Geithner_2006,Burigo_2019,Gom__2013,Sa_e_2024}.
However, the standard $S_\mathrm{N}$ data supporting this approximation are calculations of electromagnetic interactions only \citep{ICRU_2016}, and the significance of hadronic interactions has not been assessed.

\subsubsection{Electronic stopping power}
Electronic stopping power $S_\mathrm{e}$, also known as unrestricted linear energy transfer $L = L_\mathit{\Delta \to \infty}$, is the mean energy transferred from a charged particle to atomic electrons in traversing a unit distance in matter \citep{ICRU_2016}.
For relativistic protons and ions, the Bethe theory \citep{Bethe_1953} with the density-effect correction $\delta$ \citep{Sternheimer_1984} gives
\begin{equation}
S_\mathrm{e} = 4 \text{\pi}\, r_\mathrm{e}^2\, m_\mathrm{e} c^2\, n_\mathrm{e}\, \frac{z^2}{\beta^2} \left(
\ln \frac{2 m_\mathrm{e} c^2}{I} + \ln \frac{\beta^2}{1-\beta^2} -\beta^2
 - \frac{\delta(\beta)}{2} \right),
\label{eq_bethe}
\end{equation}
where the classical electron radius $r_\mathrm{e}$ and the electron rest energy $m_\mathrm{e} c^2$ are physical constants, the electron density $n_\mathrm{e}$ and the mean excitation energy $I$ are material properties, and the charge $z e$ and the speed $\beta c$ are particle properties.
The density effect $\delta(\beta)$ is a function specific to the material and only works for highly relativistic particles in dense matter, specifically at $\beta > 0.8667$ in water \citep{Sternheimer_1984} corresponding to the kinetic energy $E > \qty{0.513}{MeV}$ for electrons and the specific kinetic energy $E/m > \qty{936}{MeV/u}$ for particles of mass $m$.
The higher-order corrections to $S_\mathrm{e}$ are negligible for ionizing particles with $\beta \gtrsim 0.2$ in reference dosimetry conditions as verified in \ref{app_B}.

\subsubsection{Chamber perturbation}\label{sec_ch}
While secondary electrons produced by ionization mostly stop immediately, some energetic ones may escape from the cavity and violate the Bragg--Gray condition.
These \delta-ray electrons are incorporated into the beam to modify its quality $Q$, and $(s_\mathrm{w/air}\,p_\mathrm{ch})_Q = (D/D_\mathrm{ch})_Q$ is determined directly by detailed Monte Carlo simulation \citep{Andreo_2018,Wulff_2018,Khan_2021}, where ${D_\mathrm{ch}}_Q = \sum_{i \in Q} \int_{\ell \in V_\mathrm{ch}} ({S_\mathit{\Delta\,}}_i/\rho)_\mathrm{air} \mathrm{d}\ell / V_\mathrm{ch}$ is the dose to air in the cavity volume $V_\mathrm{ch}$, $S_\mathit{\Delta} = S_\mathrm{N} + L_\mathit{\Delta} \approx L_\mathit{\Delta\,}$ is the stopping power with individual energy transfer to an atomic electron restricted up to the \delta-ray production cutoff $\mathit{\Delta}$ \citep{Spencer_1955}, and $\ell$ is the track of charged particle $i$ in the beam. 
The cutoff energy $\mathit{\Delta}$ is chosen such that secondary electrons of $E < \mathit{\Delta}$ can be considered localized within the cavity.
The dose to water in a small volume, $D_Q$, is similarly determined with the ionization chamber virtually removed.
Since the Monte Carlo simulation stochastically deals with complex particle interactions, the resulting $(D/D_\mathrm{ch})_Q$ accounts for the realistic effects of an ionization chamber apart from the Bragg--Gray cavity. 

The chamber perturbation factor $p_\mathrm{ch}$ can be extracted from $(s_\mathrm{w/air}\,p_\mathrm{ch})_Q$ as ${p_\mathrm{ch}}_Q = (s_\mathrm{w/air}\,p_\mathrm{ch})_Q/{s_\mathrm{w/air}}_Q$.
 In ICP, the mean water-to-air mass stopping power ratio ${s_\mathrm{w/air}}_Q = \sum_{i \in Q} {s_\mathrm{w}}_i/\sum_{i \in Q} {s_\mathrm{air}}_i$ is unrestricted for ion beams or restricted for the other beam types with a cut-off energy $\mathit{\Delta}=\qty{10}{keV}$ or $\beta \lesssim 0.2$ for secondary electrons \citep{IAEA_2024}. 
The restricted ${s_\mathrm{w/air}}_Q$ involves some cavity-size effect which interferes with ${p_\mathrm{ch}}_Q$.

\subsection{New conceptualization and formulation}

As discussed in \ref{sec_conventional}, nuclear interactions are occasionally disregarded or inconsistently handled in dosimetry. 
Although their effects are negligible for conventional photon and electron beams, they might contribute to inaccuracies for proton and ion beams.
According to global consensus reports, all subsequent chemical reactions and biological damages, including DNA strand breaks, originate from atomic excitation and ionization \citep{UNSCEAR_2000,BEIR_VII_2006}.
It may thus be reasonable to deliberately exclude nuclear energies from medical dosimetry. 
Based on this idea, we will develop a new dosimetry framework that exclusively counts electronic energy deposits.

\subsubsection{Electronic $W$ value}

The secondary electrons produced by ionization may recursively cause electronic collisions to form ionization clusters. 
In the standard theory of $W$ value \citep{ICRU_1979}, the total energy deposited in electronic collisions is modeled as
\begin{equation}
E_\mathrm{e} = N_\mathrm{i} (\overline{E_\mathrm{i}} + \overline{\epsilon}) + N_\mathrm{x} \, \overline{E_\mathrm{x}},
\end{equation}
where $N_\mathrm{i}$ is the total number of electrons produced by ionization, $\overline{E_\mathrm{i}}$ is the mean energy expended to produce an electron, $\overline{\epsilon}$ is the mean energy of electrons too slow to excite or ionize a molecule, $N_\mathrm{x}$ is the total number of discrete excited states produced, and $\overline{E_\mathrm{x}}$ is the mean energy of discrete excited states.
The mean energy expended by electronic collisions in matter per ion pair formed, or the electronic $W$ value,
\begin{equation}
W_\mathrm{e} = \frac{E_\mathrm{e}}{N_\mathrm{i}} = \overline{E_\mathrm{i}} + \overline{\epsilon} + \frac{N_\mathrm{x}}{N_\mathrm{i}} \, \overline{E_\mathrm{x}},
\end{equation}
is assumed to be a material-specific constant as well as $\overline{E_\mathrm{i}}$, $\overline{E_\mathrm{x}}$, $\overline{\epsilon}$, and $N_\mathrm{x}/N_\mathrm{i}$ asymptotically at $E_\mathrm{e} \gg \overline{\epsilon}$, which generally applies to ionizing radiations.
The constant $N_\mathrm{x}/N_\mathrm{i}$, or the proportionality between atomic excitation and ionization, implies that ionization will serve as a surrogate for both and subsequent chemical and biological effects. 

\subsubsection{Electronic dose}

We define a new quantity, electronic dose, as the radiant energy imparted to a unit mass of matter by electronic collisions.
For photon and electron beams, it is equivalent to the absorbed dose as their energy deposits are solely by energetic electrons colliding with atomic electrons.
In analogy with Eqs.~(\ref{eq_D})--(\ref{eq_kQ}), ionization chambers can measure the electronic dose to water by
\begin{equation}
D_\mathrm{e} = \frac{M}{e}\,\frac{{W_\mathrm{e}}_\mathrm{air}}{\rho_\mathrm{air}\,V_\mathrm{ch}}\,{{s_\mathrm{e}}_\mathrm{w/air}}_Q\,{p_\mathrm{ch}}_Q 
=M \, {N_\mathrm{D}}_{Q_0} \, ({s_\mathrm{e}}_\mathrm{w/air} \, p_\mathrm{ch})_{Q/Q_0},
\label{eq_De}
\end{equation}
where ${s_\mathrm{e}}_\mathrm{w/air} = (S_\mathrm{e}/\rho)_\mathrm{w/air}$ is the water-to-air mass electronic stopping power ratio and ${N_\mathrm{D}}_{Q_0} = (D/M)_{Q_0} = (D_\mathrm{e}/M)_{Q_0}$ is the dosimeter calibration coefficient for $^{60}$Co \gamma\ rays.

With the approximation $s_\mathrm{w/air} \approx {s_\mathrm{e}}_\mathrm{w/air}$ verified in \ref{app_B}, the beam-quality correction factor in Eq.~(\ref{eq_De}) for standard dosimeter models is derived retrospectively by,
\begin{equation}
({s_\mathrm{e}}_\mathrm{w/air}\,p_\mathrm{ch})_{Q/Q_0} \approx (s_\mathrm{w/air}\,p_\mathrm{ch})_{Q/Q_0} = k_{Q/Q_0} \frac{{W_\mathrm{air}}_{Q_0}}{{W_\mathrm{air}}_Q},
\label{eq_spQ}
\end{equation}
from the $k_{Q/Q_0}$ and ${W_\mathrm{air}}_Q$ data specified in Table~\ref{tab_parameters}.
This derivation excludes the ${W_\mathrm{air}}_{Q/Q_0}$ factor and its uncertainty contribution from the $k_{Q/Q_0}$.
The relative standard uncertainty in the determination of electronic dose is estimated by
\begin{equation}
\hat\Delta D_\mathrm{e} = \sqrt{(\hat\Delta M)^2 + (\hat\Delta {N_\mathrm{D}}_{Q_0})^2 + (\hat\Delta {s_\mathrm{w/air}}_{Q/Q_0})^2+ (\hat\Delta {p_\mathrm{ch}}_{Q/Q_0})^2}.
\label{eq_DeltaDe}
\end{equation}

\begin{table}[htb]
\caption{Dosimetric parameters compiled in the international code of practice for standard ionization chambers \protect\citep{IAEA_2024}.}
\label{tab_parameters}
\begin{tabular}{lccc}
\hline\hline
Beam type & $k_{Q/Q_0}$ & ${W_\mathrm{air}}_Q$ & ${s_\mathrm{w/air}}_{Q}$\mpfootnotemark[1] \\
\hline
$^{60}$Co \gamma\ & 1 & \qty{33.97}{eV} & \num{1.127} \\
Photon & Table 16 & \qty{33.97}{eV} & --- \\
Electron &Table~20 & \qty{33.97}{eV} & --- \\
Proton & Table~37 & \qty{34.44}{eV} & Eq.~(100)\mpfootnotemark[2] \\
Ion & Table~42 & \qty{34.71}{eV} & \num{1.126} \\
\hline\hline
\end{tabular}
\footnotetext[1]{Except for ion beams, $s_\mathrm{w/air}$ is restricted with $\mathit{\Delta}=\qty{10}{keV}$.}
\footnotetext[2]{$s_\mathrm{w/air} (R_\mathrm{res}) = 1.131 - \num{2.327e-5} R_\mathrm{res}/\unit{cm} + \qty{2.046e-3}{cm}/R_\mathrm{res}$.}
\end{table}

\subsubsection{Water-equivalent gas ionometric dosimetry} \label{sec_weg}

The ionometric dosimetry of electronic dose can be extended to non-air gas chambers, which may be used in waterproof ionization chambers customized with gas sealing.
For the gas, water equivalence is evaluated by the insensitivity of the correction factor ${s_\mathrm{e}}_\mathrm{w/gas}$ to the particle properties.
In the ratio ${s_\mathrm{e}}_\mathrm{w/gas} = (S_\mathrm{e}/\rho)_\mathrm{w}/(S_\mathrm{e}/\rho)_\mathrm{gas}$ using Eq.~(\ref{eq_bethe}), the particle $\beta$ dependence generally exists due to the density-effect terms ($\delta_\mathrm{w} \neq \delta_\mathrm{gas}$).
However, since the density effect is absent in water as well as in gas ($\delta_\mathrm{w}=\delta_\mathrm{gas}=0$) for therapeutic protons and ions of $E/m \leq \qty{430}{MeV/u}$ \citep{IAEA_2024}, the $\beta$ dependence cancels out for $I _\mathrm{gas} = I _\mathrm{w}$.

Water-equivalent gas (WEG) for proton and ion beams is defined to have $I_\mathrm{weg} = I_\mathrm{w}$, and the water-to-WEG mass electronic stopping power ratio will be a gas-specific constant,
\begin{equation}
 {{s_\mathrm{e}}_\mathrm{w/weg}}_Q = (Z/A_\mathrm{r})_\mathrm{w/weg}, 
 \label{eq_sweg}
\end{equation}
where $Z$ and $A_\mathrm{r}$ are the mean values of atomic number and relative atomic mass, and $(Z/A_\mathrm{r}) =n_\mathrm{e} \unit{u}/\rho$ is the relative electron content of the material.
Nitrogen-based WEGs have been designed in \ref{app_A} and validated in \ref{app_B}.
As shown in \ref{app_C}, they are also valid for $^{60}$Co \gamma\ rays, and then Eq.~(\ref{eq_sweg}) also applies to $Q_0$.
With the cancellation of ${{s_\mathrm{e}}_\mathrm{w/weg}}_{Q/Q_0}$, Eq.~(\ref{eq_De}) is modified to 
\begin{equation}
D_\mathrm{e} = M\,{N_\mathrm{D}}_{Q_0}\,{p_\mathrm{ch}}_{Q/Q_0}, 
\label{eq_Deweg}
\end{equation}
for calibrated WEG ionization chambers.

A finite $I_\mathrm{weg}$ error with respect to $I_\mathrm{w}$ will contribute to the measurement error of $D_\mathrm{e}$ through propagation.
Using the relation $D_\mathrm{e} \propto M \propto S_\mathrm{e}$ and Eq.~(\ref{eq_bethe}) with $\delta(\beta) = 0$, the relative standard uncertainty of $D_\mathrm{e}$ due to $I_\mathrm{weg} \neq I_\mathrm{w}$ will be
\begin{equation}
\begin{split}
& \hat\Delta {D_\mathrm{e}}_I = \left| \frac{\partial D_\mathrm{e}}{D_\mathrm{e} \partial I} \right| \hat\Delta (I_\mathrm{weg}-I_\mathrm{w}) \\
& = \left( \ln \frac{2 m_\mathrm{e} c^2}{I_\mathrm{w}} + \ln \frac{\beta^2}{1-\beta^2} - \beta^2 \right)^{-1} 
\sqrt{(\hat\Delta I_\mathrm{weg})^2+(\hat\Delta I_\mathrm{w})^2}.
\end{split}
\end{equation}
With $\Delta I_\mathrm{weg} = \qty{1.1}{eV}$ for the WEGs designed in \ref{app_A} with a standard inaccuracy of $\hat\Delta v_\mathrm{L} = \qty{5}{\%}$ for gas mixing, this amounts to $\hat\Delta {D_\mathrm{e}}_I \leq \qty{0.36}{\%}$ at $\beta \geq 0.2$ or $E/m \geq \qty{20}{MeV/u}$ for protons and ions corresponding to residual range $R_\mathrm{res} \geq \qty{0.5}{cm}$ in water for reference dosimetry \citep{IAEA_2024}.
Apart from the $\hat\Delta {D_\mathrm{e}}_I = \qty{0.36}{\%}$ propagated through $\hat\Delta I_\mathrm{weg}$, the inaccuracy $\hat\Delta v_\mathrm{L}$ will not affect $D_\mathrm{e}$ measurement, provided that gas mixing is precisely consistent between dosimeter calibration and dose measurement.

The remaining beam-quality correction factor ${p_\mathrm{ch}}_{Q/Q_0}$ in Eq.~(\ref{eq_Deweg}) is to correct overall dosimetric effects originated from the chamber structure, and then the value for the same dosimeter model operated with air should be applicable.
The ${p_\mathrm{ch}}_{Q/Q_0}$ values for standard ionization chambers are unfortunately not directly available in ICP, but can be derived retrospectively by
\begin{equation}
{p_\mathrm{ch}}_{Q/Q_0} = k_{Q/Q_0}\,\frac{{W_\mathrm{air}}_{Q_0}}{{W_\mathrm{air}}_Q}\,\frac{{s_\mathrm{w/air}}_{Q_0}}{{s_\mathrm{w/air}}_Q},
\label{eq_pQ}
\end{equation}
from the $k_{Q/Q_0}$, ${W_\mathrm{air}}_Q$, and ${s_\mathrm{w/air}}_Q$ data specified in Table~\ref{tab_parameters}.
This derivation excludes the ${W_\mathrm{air}}_{Q/Q_0}$ and ${s_\mathrm{w/air}}_{Q/Q_0}$ factors and their uncertainty contributions from the $k_{Q/Q_0}$.

ICP evaluated unrestricted $s_\mathrm{w/air}$ for ion beams and restricted $s_\mathrm{w/air}$ with $\mathit{\Delta}=\qty{10}{keV}$ for other beams \citep{IAEA_2024}, which will introduce some additional uncertainty.
With the restriction, \citet{Nahum_1978} found \qty{0.5}{\%} change for $^{60}$Co \gamma\ rays, and \citet{Medin_1997} found \qty{0.1}{\%} to \qty{0.6}{\%} changes for proton beams.
Based on these facts, we estimate the relative uncertainty of $D_\mathrm{e}$ originated from the stopping power restriction in ICP to be $\hat\Delta {D_\mathrm{e}}_\mathit{\Delta} = \qty{0.5}{\%}$.
Consequently, the relative standard uncertainty in the electronic-dose determination is estimated by
\begin{equation}
\begin{split}
\hat\Delta D_\mathrm{e} &= \Bigl[ (\hat\Delta M)^2 + (\hat\Delta {N_\mathrm{D}}_{Q_0})^2 + (\hat\Delta {p_\mathrm{ch}}_{Q/Q_0})^2 \\
&+ (\hat\Delta {D_\mathrm{e}}_I)^2 + (\hat\Delta {D_\mathrm{e}}_\mathit{\Delta})^2 \Bigr]^\frac{1}{2}.
\label{eq_DeltaDeweg}
\end{split}
\end{equation}

\subsection{Thought reference dosimetry}

The proposed reference dosimetry was tested in a thought experiment using two Farmer-type ionization chambers of one model (Type 30013, PTW, Freiburg, Germany), one operated with air and the other with WEG.
Note that the whole procedure here was manipulation of published numbers and easily applicable to the other 18 cylindrical and 9 plane-parallel models covered by ICP.

The two ionization chambers were assumed to have been calibrated with $^{60}$Co \gamma\ rays to determine ${N_\mathrm{D}}_{Q_0}$ for each.
Table~\ref{tab_beams} shows the test beams used and their $k_{Q/Q_0}$ values taken from ICP, where the air chamber was used for all the test beams while the WEG chamber was used for the proton and ion beams only.
For all the beams, hypothetical measurement was conducted at a reference depth to determine the doses for a fixed ionization charge, $M = \qty{1}{Gy}/{N_\mathrm{D}}_{Q_0}$, namely the ionization for \qty{1}{Gy} from $^{60}$Co \gamma\ rays, for convenient comparison around $\qty{1}{Gy}$.
The $M$ was converted to $D$ by Eq.~(\ref{eq_Dref}) and to $D_\mathrm{e}$ by Eq.~(\ref{eq_De}) for the air chamber, and to $D_\mathrm{e}$ by Eq.~(\ref{eq_Deweg}) for the WEG chamber. 
Their relative standard uncertainties were evaluated by Eqs.~(\ref{eq_DeltaD}), (\ref{eq_DeltaDe}), and (\ref{eq_DeltaDeweg}), respectively.

\begin{table}[htb]
\caption{Test beams and their dosimetric parameters (reference depth in water $z_\mathrm{ref}$, tissue-phantom ratio at \qty{20}{cm} TPR$_{20/10}$, median range $R_\mathrm{50}$, residual range $R_\mathrm{res}$, beam-quality correction factor $k_{Q/Q_0}$) in thought reference dosimetry with a Farmer-type ionization chamber (PTW Type 30013), based on the international code of practice \protect\citep{IAEA_2024}.}
\label{tab_beams}
\begin{tabular}{llclc}
\hline\hline
Beam type & Energy & $z_\mathrm{ref}$ & Quality index & $k_{Q/Q_0}$ \\
\hline
Photon & \qty{6}{MV} & \qty{10}{cm} & $\mathrm{TPR}_{20/10} = 0.68$ & \num{0.9876} \\
Electron & \qty{12}{MeV} & \qty{2.9}{cm} & $R_{50} = \qty{5.0}{cm}$ & \num{0.9155} \\
Proton & \qty{156}{MeV} & \qty{2.0}{cm} & $R_\mathrm{res} = \qty{15.0}{cm}$ & \num{1.0245} \\
Carbon ion & \qty{290}{MeV/u} & \qty{2.0}{cm} & \multicolumn{1}{c}{---} & \num{1.028}\phantom{0} \\
\hline\hline
\end{tabular}
\end{table}

\section{Results}

Table~\ref{tab_dosimetry} shows the results of the thought reference dosimetry.
With the fixed $M = \qty{1}{Gy}/{N_\mathrm{D}}_{Q_0}$, the $D/\unit{Gy}$ values coincided with their respective $k_{Q/Q_0}$ values.
For photon and electron beams, $D_\mathrm{e}$ was shown to be identical to $D$.
For proton and ion beams, the $D_\mathrm{e}$ by the air chamber was lower than $D$ due to the exclusion of ${W_\mathrm{air}}_{Q/Q_0}$.
The WEG chamber dosimetry further excluded the ${{s_\mathrm{e}}_\mathrm{w/air}}_{Q/Q_0}$ correction, and the $D_\mathrm{e}/\unit{Gy}$ values coincided with their respective ${p_\mathrm{ch}}_{Q/Q_0}$ values.
The exclusion of these constituent correction factors contributed to the reduction of the dosimetric uncertainty as well as the correction itself.

\begin{table}[htb]
\caption{Results of thought reference dosimetry with air and water-equivalent gas (WEG) ionization chambers: absorbed dose $D$ and electronic dose $D_\mathrm{e}$ at a fixed ionization charge for \qty{1}{Gy} of $^{60}$Co \gamma\ rays with their relative standard uncertainty in parentheses.}
\label{tab_dosimetry}
\begin{tabular}{lccc}
\hline\hline
& \multicolumn{2}{c}{Air} & WEG \\
\cline{2-3} 
Beam type & $D/\unit{Gy}$ & $D_\mathrm{e}/\unit{Gy}$ & $D_\mathrm{e}/\unit{Gy}$ \\
\hline
Photon & \num{0.988} (\qty{1.0}{\%}) & \num{0.988} (\qty{1.0}{\%}) & \multicolumn{1}{c}{---} \\
Electron & \num{0.916} (\qty{1.1}{\%}) & \num{0.916} (\qty{1.1}{\%}) & \multicolumn{1}{c}{---} \\
Proton & \num{1.025} (\qty{1.6}{\%}) & \num{1.011} (\qty{1.5}{\%}) & \num{1.007} (\qty{1.2}{\%}) \\
Carbon ion & \num{1.028} (\qty{2.6}{\%}) & \num{1.006} (\qty{2.1}{\%}) & \num{1.007} (\qty{1.5}{\%}) \\
\hline\hline
\end{tabular}
\end{table}

\section{Discussion}

For high-energy photon and electron beams with standard ionization chambers, electronic dose is equivalent to absorbed dose, and the dosimetric uncertainty, \qty{1.0}{\%} and \qty{1.1}{\%}, may already be sufficiently small.
For proton and ion beams with WEG ionization chambers, the expected uncertainty of electronic dose, \qty{1.2}{\%} and \qty{1.5}{\%}, is comparable to the dosimetric uncertainty for photon and electron beams.
On clinical transition from absorbed dose to electronic dose, the change in apparent dose value by \(-1.4\%\) for proton beams and \(-2.2\%\) for ion beams must be considered. 
Physicians may either reduce the dose value to prescribe accordingly or accommodate the increase in actual dosing.
This decision-making process will be analogous to the instances of treatment system upgrades.
In proton and ion beam therapy, clinical dose prescription generally involves the relative biological effectiveness (RBE) of the radiation \citep{ICRU_2007,ICRU_2019}.
The RBE models are continuously developing to be eventually integrated into clinical systems \citep{Friedrich_2026}.

In the thought reference dosimetry, we derived the beam-quality correction factors, $(s_\mathrm{w/air}\,p_\mathrm{ch})_{Q/Q_0}$ and ${p_\mathrm{ch}}_{Q/Q_0}$, retrospectively from the $k_{Q/Q_0}$ and other parameters in ICP for proof of concept.
To formally establish new dosimetry protocols for electronic dose, it is more natural and potentially more accurate to newly determine these correction factors by detailed Monte Carlo simulation per beam quality per dosimeter model as done for absorbed dose \citep{Kawrakow_2000,Kretschmer_2020,Urago_2023,Urago_2025,Stolzenberg_2025}, where
the ratio of electronic-dose-to-water to electronic-dose-to-chamber, $D_\mathrm{e}/{D_\mathrm{e}}_\mathrm{ch}$, should be determined by collecting electronic energy deposits in analogy to $D/D_\mathrm{ch}$ described in \ref{sec_ch}.

This study was aimed at improving the accuracy of beam-quality correction in ionometric dosimetry.
Alternatively, calorimetric absolute dosimetry for both reference and clinical beams can directly determine the $k_{Q/Q_0}$ factor, not depending on standard data nor Monte Carlo simulation and free from their uncertainty.
In fact, \citet{Osinga_Bl_ttermann_2017} and \citet{Holm_2021,Holm_2022} experimentally determined the $k_{Q/Q_0}$ factors for carbon-ion beams at a small uncertainty of \qty{0.8}{\%}.
That should be compared to the uncertainty in beam-quality correction for electronic dose, $\surd[(\hat\Delta {{s_\mathrm{e}}_\mathrm{w/air}}_{Q/Q_0})^2 + (\hat\Delta {p_\mathrm{ch}}_{Q/Q_0})^2] = \qty{1.9}{\%}$ with a standard ionization chamber and $\surd[(\hat\Delta {p_\mathrm{ch}}_{Q/Q_0})^2 + (\hat\Delta {D_\mathrm{e}}_I)^2 + (\hat\Delta {D_\mathrm{e}}_\mathit{\Delta})^2] = \qty{1.2}{\%}$ with a WEG ionization chamber.
Nevertheless, such experimental determination may be challenging for standardization with a variety of dosimeter models and beam qualities.
In contrast, the proposed dosimetry framework for electronic dose will perfectly align with the international radiation dosimetry system for traceable dosimeter calibration, where calorimetry provides reference absolute doses of $^{60}$Co \gamma\ rays of $E \lesssim \qty{1.33}{MeV}$ below the photo-nuclear threshold at primary standards calibration laboratories.

This conceptual study is based on the consensus assumption that chemical and biological effects of radiation originate from electronic collisions.
In a soft nuclear collision, the energy transferred to a nucleus converts to molecular motion including vibration and rotation, and dissipates into heat.
In a hard nuclear collision, a recoil nucleus or atom emits from the molecule and joins the radiation as a secondary particle.
In other words, nuclear collisions contribute little to chemical reactions except for initial molecular transmutation.
The initial damage has been evaluated extensively for crystalline materials, but not for biological materials \citep{Nordlund_2018}.

Another key assumption is that the $W_\mathrm{e}$ value is a material-specific constant, or that the radiation dependence of the $W$ value is attributed to nuclear stopping, which naturally depends on the projectile particle mass and energy. 
In recent studies similarly based on the standard $W$ value theory \citep{ICRU_1979}, the same assumption was employed for ionization quenching factor $W_\mathrm{e}/W$ \citep{Katsioulas_2022,Foresi_2026}, suggesting the validity for the assumption.
In nuclear collisions, while absorbed dose is designed to exclude heat defect for target nuclear transmutation, electronic dose additionally excludes residual kinetic energies of recoil nuclei below atomic excitation energies to dissipate as heat without ionization. 
This exclusion accounts for the asymptotic invariance of the $W_\mathrm{e}$ value for ionizing radiations in contrast to the radiation-dependent $W$ value.

The concept of stopping evaluates the energy lost by particles and does not exactly apply to dose, which evaluates the energy absorbed by matter, in the presence of extra energy dissipation or outflow. 
This conceptual misapplication may historically be attributed to the Bragg--Gray cavity theory. 
With the synonymity between $L$ and $S_\mathrm{e}$, it may be reasonable to take mass linear energy transfer $l=L/\rho$ instead of mass electronic stopping power $s_\mathrm{e} = S_\mathrm{e}/\rho$ in the formulation of electronic dose $D_\mathrm{e}$ in Eq.~(\ref{eq_De}). 

\section{Conclusions}

We have conceptualized an alternative quantity to measure radiotherapy doses, electronic dose, which is expected to enhance theoretical consistency and medical relevance.
The formulated framework will be applicable to radiotherapy dosimetry with traceable dosimeter calibration under the international radiation dosimetry system.

With standard ionization chambers, electronic dose is equivalent to absorbed dose for photon and electron beams.
For proton and ion beams, the transition from absorbed dose to electronic dose will change apparent dose values by \qty{-1.4}{\%} and \qty{-2.2}{\%}, and decrease the uncertainty from \qty{1.6}{\%} to \qty{1.5}{\%} and from \qty{2.6}{\%} to \qty{2.1}{\%}, respectively.

Nitrogen-based WEG mixtures for proton and ion beams have been designed for gas-sealed ionization chambers, which will further decrease the uncertainty to \qty{1.2}{\%} for proton beams and to \qty{1.5}{\%} for ion beams. 
 
\section*{Acknowledgment}

This work was supported by JSPS KAKENHI Grant Number JP26K10623.

\appendix

\section{Water-equivalent gas mixtures}\label{app_A}

\setcounter{equation}{0}
\setcounter{table}{0}
\setcounter{figure}{0}

We consider mixing two gases of high and low $I$ values to achieve the same $I$ value as water, $I_\mathrm{weg} = I_\mathrm{w}$.
As shown in Table~\ref{tab_gases}, we chose nitrogen, which is a main component of air with an $I$ value slightly higher than water, for a base gas, and methane and ethane, which are standard quenching gases for proportional counters with similar low $I$ values, for additive gas candidates.
These WEG mixtures are expected to minimize ion recombination by eliminating electronegative gas components, such as oxygen, which would absorb free electrons.
In air ionization, negatively charged oxygen molecular ions drift slowly and tend to recombine with positive ions, especially when they pile up at high dose rates.

\begin{table}[htb]
\caption{Properties of water and relevant gases: density $\rho$ (\qty{20}{\degreeCelsius}, \qty{101.325}{kPa}), relative electron content $(Z/A_\mathrm{r})$, and mean excitation energy $I$ with standard uncertainty $\Delta I$ \protect\citep{ICRU_1984,ICRU_2016}.} 
\label{tab_gases}
\begin{tabular}{lcccc}
\hline\hline
Material \rule{0pt}{1.5em} & $\dfrac{\rho}{\unit{kg/m^3}}$ & $(Z/A_\mathrm{r})$ & $\dfrac{I}{\unit{eV}}$ & $\dfrac{\Delta I}{\unit{eV}}$ \\ 
\hline
Water & \num{998.23} & \num{0.55509} & \num{78}\mpfootnotemark[1] & \num{2}\mpfootnotemark[1] \\
Air & \num{1.2048} & \num{0.49919} & \num{85.7} & \num{1.2}\mpfootnotemark[1] \\
Nitrogen & \num{1.1653} & \num{0.49976} & \num{82.3}\mpfootnotemark[1] & \num{1.2}\mpfootnotemark[1] \\
Methane & \num{0.6672} & \num{0.62334} & \num{41.7} & \num{0.8} \\
Ethane & \num{1.2630} & \num{0.59861} & \num{45.4} & \num{0.9} \\
\hline\hline
\end{tabular}
\footnotetext[1]{Updated in \citet{ICRU_2016}.}
\end{table}

For the mixing of low-$I$ gas (L) and high-$I$ gas (H) by mass fractions $w_\mathrm{L}$ and $w_\mathrm{H}=1-w_\mathrm{L}$, the Bragg additivity rule gives the relations of gas properties among L, H, and WEG \citep{ICRU_1984},
\begin{gather}
1/\rho_\mathrm{weg} = w_\mathrm{L}/\rho_\mathrm{L} + w_\mathrm{H}/\rho_\mathrm{H}, \label{eq_rho} \\
(Z/A_\mathrm{r})_\mathrm{weg} = w_\mathrm{L}\,(Z/A_\mathrm{r})_\mathrm{L} + w_\mathrm{H}\,(Z/A_\mathrm{r})_\mathrm{H}, \label{eq_Z}\\
(Z/A_\mathrm{r})_\mathrm{weg}\,\ln I_\mathrm{weg} = w_\mathrm{L}\,(Z/A_\mathrm{r})_\mathrm{L}\,\ln I_\mathrm{L} + w_\mathrm{H}\,(Z/A_\mathrm{r})_\mathrm{H}\,\ln I_\mathrm{H}. \label{eq_ZI}
\end{gather}
With $I_\mathrm{weg} = I_\mathrm{w}$ by definition, the weight fraction is solved as
\begin{equation}
w_\mathrm{L} = \frac{(Z/A_\mathrm{r})_\mathrm{H}\,\ln I_\mathrm{H/w}}{(Z/A_\mathrm{r})_\mathrm{H}\,\ln I_\mathrm{H/w}-(Z/A_\mathrm{r})_\mathrm{L}\,\ln I_\mathrm{L/w}},
\end{equation}
to give $v_\mathrm{L} = \rho_\mathrm{weg/L} \,w_\mathrm{L}$ in volume fraction or mole fraction.
With independent uncertainties of $I_\mathrm{L}$, $I_\mathrm{H}$, and $w_\mathrm{L}$, the relative standard uncertainty of $I_\mathrm{weg}$ in Eq.~(\ref{eq_ZI}) is given by
\begin{equation}
\begin{split} 
&\hat\Delta I_\mathrm{weg} = \frac{\Delta I_\mathrm{weg}}{I_\mathrm{weg}} = \Delta \ln I_\mathrm{weg} \\
&=\sqrt{\left(\frac{\partial \ln I_\mathrm{weg}}{\partial I_\mathrm{L}} \Delta I_\mathrm{L}\right)^2 + \left(\frac{\partial \ln I_\mathrm{weg}}{\partial I_\mathrm{H}} \Delta I_\mathrm{H}\right)^2 + \left(\frac{\partial \ln I_\mathrm{weg}}{\partial w_\mathrm{L}} \Delta w_\mathrm{L}\right)^2 } \\
&=\bigg\{ \left[w_\mathrm{L}\,(Z/A_\mathrm{r})_\mathrm{L/weg}\,\hat\Delta I_\mathrm{L}\right]^2 
+ \left[w_\mathrm{H}\,(Z/A_\mathrm{r})_\mathrm{H/weg}\,\hat\Delta I_\mathrm{H}\right]^2 \\
&+\Bigl[ w_\mathrm{L}\,(Z/A_\mathrm{r})_\mathrm{L/H} + w_\mathrm{H} \Bigr]^{-4} \left[\mathstrut w_\mathrm{L}\,(Z/A_\mathrm{r})_\mathrm{L/H}\,\ln I_\mathrm{L/H}\,\hat\Delta w_\mathrm{L} \right]^2 \bigg\}^{\frac{1}{2}}.
\label{eq_DelI}
\end{split}
\end{equation}

By applying the material properties in Table~\ref{tab_gases} to Eqs.~(\ref{eq_rho})--(\ref{eq_DelI}) with an assumption of $\hat\Delta w_\mathrm{L} = \hat\Delta v_\mathrm{L} = \qty{5}{\%}$ as a standard manufacturing accuracy for gas mixing, the properties of the WEGs were determined as shown in Table~\ref{tab_weg}.

\begin{table}[htb]
\caption{Properties of nitrogen-based water-equivalent gas mixtures for proton and ion beams: mass and volume fractions $w_\mathrm{L}$ and $v_\mathrm{L}$ of low-$I$ additive gas, density $\rho$ (\qty{20}{\degreeCelsius}, \qty{101.325}{kPa}), relative electron content $(Z/A_\mathrm{r})$, and mean excitation energy $I$ with standard uncertainty $\Delta I$.}
\label{tab_weg}
\begin{tabular}{lcccccc}
\hline\hline
Additive gas \rule{0pt}{1.5em} & $\dfrac{w_\mathrm{L}}{\unit{w t\%}}$ & $\dfrac{v_\mathrm{L}}{\unit{vol \%}}$ & $\dfrac{\rho}{\unit{kg/m^3}}$ & $(Z/A_\mathrm{r})$ & $\dfrac{I}{\unit{eV}}$ & $\dfrac{\Delta I}{\unit{eV}}$ \\
\hline
Methane & \num{6.429} & \num{10.71} & \num{1.1119} & \num{0.50770} & \num{78.0} & \num{1.1} \\
Ethane & \num{7.645} & \num{7.096} & \num{1.1722} & \num{0.50732} & \num{78.0} & \num{1.1} \\
\hline\hline
\end{tabular}
\end{table}

\section{Validation of water equivalence}\label{app_B}
\setcounter{equation}{0}
\setcounter{table}{0}
\setcounter{figure}{0}

We here verify the water equivalence of a WEG mixture, using material databases compiled by the National Institute of Standards and Technology (NIST).
The NIST PSTAR and ASTAR databases give the electronic and nuclear mass stopping powers and the continuous-slowing-down-approximation and projected ranges for protons and \alpha\ particles \citep{Berger_2017}.
The electronic stopping power included the Bethe term, the Barkas, Bloch, and shell corrections, and fitting to experimental data at low energies.
The nuclear stopping power included the elastic scattering by nuclear Coulomb potential with electronic screening, but excluded hadronic nuclear interactions.
The databases were compiled with outdated mean excitation energies \citep{ICRU_1993} used for the previous ICP \citep{IAEA_2000}, which we also use here for consistency.

\begin{figure}[htb]
\includegraphics[width=\columnwidth]{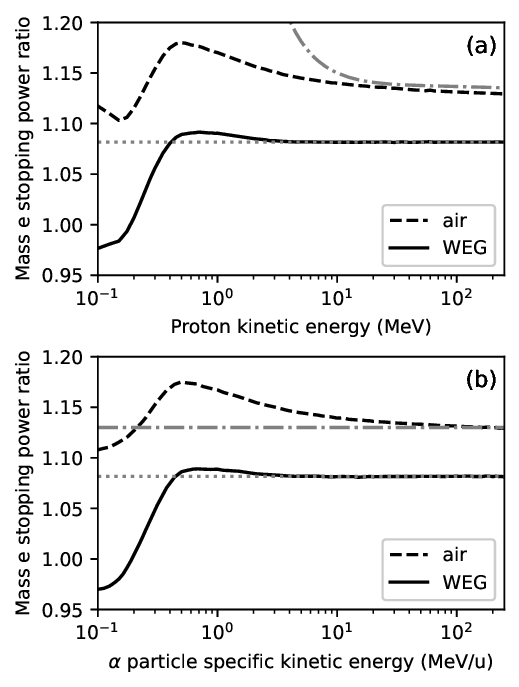}
\caption{Mass electronic stopping power ratios of water to air (dashed line) and that of water to a nitrogen--methane water-equivalent gas (WEG) (solid line) for (a) protons and (b) \alpha\ particles, based on the NIST PSTAR and ASTAR databases \citep{Berger_2017}. 
Also shown are the ICP model for air (dash-dotted lines) \citep{IAEA_2000}, and the WEG model (dotted lines).}
\label{fig_B1}
\end{figure}

Figure~\ref{fig_B1} shows the mass electronic stopping power ratio of water ($I_\mathrm{w} = \qty{75.0}{eV}$) to air ($I_\mathrm{air} = \qty{85.7}{eV}$), ${s_\mathrm{e}}_\mathrm{w/air}$, and that of water to a WEG mixture ($I_\mathrm{weg} =\qty{75.0}{eV}$) of \qty[number-unit-product=\text{-}]{10.86}{wt \%} methane ($I_\mathrm{L}=\qty{41.7}{eV}$) and \qty[number-unit-product=\text{-}]{89.14}{wt \%} nitrogen ($I_\mathrm{H}=\qty{82.0}{eV}$), ${s_\mathrm{e}}_\mathrm{w/weg}$, calculated using the NIST databases for protons ($m \approx \unit{u}$) and \alpha\ particles ($m \approx \qty{4}{u}$).
Also shown there are the $s_\mathrm{w/air}$ values recommended for beam-quality correction in the previous ICP \citep{IAEA_2000}, namely the $s_\mathrm{w/air}(R_\mathrm{res})$ function for protons, where the PSTAR projected range was applied to the residual range $R_\mathrm{res}$, and the constant $s_\mathrm{w/air} =1.13$ for ions.
For the WEG, ${s_\mathrm{e}}_\mathrm{w/weg} = (Z/A_\mathrm{r})_\mathrm{w/weg}$ is a universal constant for protons and ions.

In the energy region $E/m \gtrsim \qty{20}{MeV/u}$ or $\beta \gtrsim 0.2$ relevant to reference dosimetry at $R_\mathrm{res} \geq \qty{0.5}{cm}$, the NIST ${s_\mathrm{e}}_\mathrm{w/air}$ data were reasonably consistent with the ICP model for air, and agreed with the WEG model within \qty{0.03}{\%}.
This agreement validates the water equivalence of the WEG for reference dosimetry of electronic dose.
At low energies $E/m \lesssim \qty{0.3}{MeV/u}$, the deviations between the WEG model the NIST data were $\lesssim \qty{10}{\%}$, which should only apply to the conversion of these tiny energy deposits to contribute to dose.
While low-energy particles are involved with reference dosimetry of broad beams in a spread-out Bragg peak \citep{IAEA_2024}, legacy broad-beam delivery systems are being replaced by modern variable-energy pencil-beam scanning systems.

\begin{figure}[htb]
\includegraphics[width=\columnwidth]{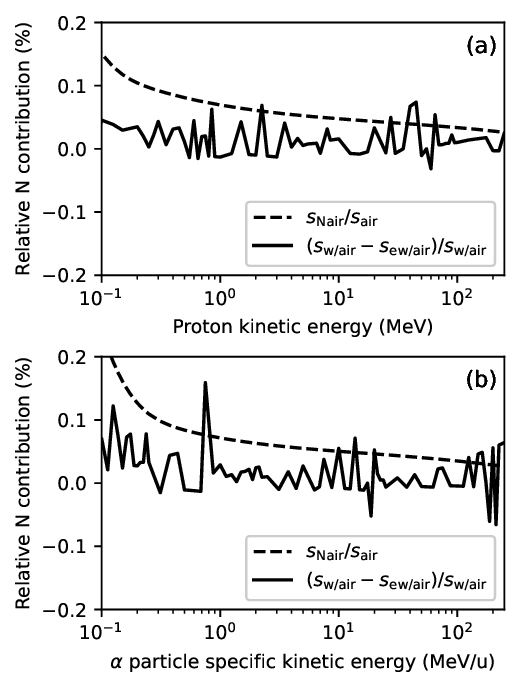}
\caption{Relative contribution of electromagnetic nuclear stopping in the mass stopping power of air ${s_\mathrm{N}}_\mathrm{air}/s_\mathrm{air}$ (dashed lines) and that in the water-to-air mass stopping power ratio $(s_\mathrm{w/air}-{s_\mathrm{e}}_\mathrm{w/air})/s_\mathrm{w/air}$ (solid lines) for (a) protons and (b) \alpha\ particles, based on the NIST PSTAR and ASTAR databases \citep{Berger_2017}.}
\label{fig_B2}
\end{figure}

As shown in Fig.~\ref{fig_B2}, the contribution of nuclear stopping in the mass stopping power of air, ${s_\mathrm{N}}_\mathrm{air}/s_\mathrm{air}$, was below \qty{0.1}{\%} for $E/m \gtrsim \qty{20}{MeV/u}$, and that in the water-to-air mass stopping power ratio, $(s_\mathrm{w/air}-{s_\mathrm{e}}_\mathrm{w/air})/s_\mathrm{w/air}$, was generally smaller but fluctuated within \qty{\pm 0.1}{\%} due to inherent limitations in the numerical precision of the NIST data. 
This confirms that electromagnetic nuclear stopping may be disregarded in reference dosimetry.

\section{The density effect for $^{60}$Co \gamma\ rays in water} \label{app_C}
\setcounter{equation}{0}
\setcounter{table}{0}
\setcounter{figure}{0}

The WEG in this study relies on the absence of the density effect in water as well as in gas. 
Regarding the reference beam quality $Q_0$ for dosimeter calibration, a $^{60}$Co isotope decays to emit two \gamma-ray photons of \qty{1.17}{MeV} and \qty{1.33}{MeV}, whose Compton edges \qty{0.96}{MeV} and \qty{1.12}{MeV} are higher than the density effect threshold \qty{0.513}{MeV} for electrons in water.
This appendix evaluates the significance of the density effect to examine the validity of the WEG for $^{60}$Co γ rays.

\begin{figure}[htb]
\includegraphics[width=\columnwidth]{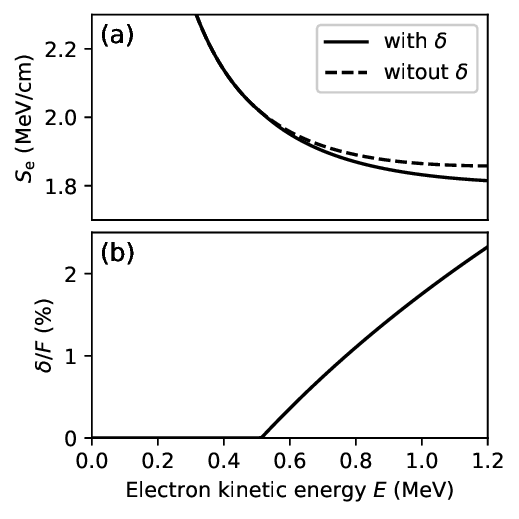}
\caption{(a) Electronic stopping power of water, $S_\mathrm{e}$, with and without the density effect $\delta$, and (b) density effect relative to main term, $\delta/F$, as a function of electron kinetic energy $E$.}
\label{fig_C1}
\end{figure}

In the Bethe formula \citep{Bethe_1953} for electronic stopping power, 
\begin{equation}
S_\mathrm{e}(\beta) = 2 \text{\pi} r_\mathrm{e}^2 m_\mathrm{e} c^2 n_\mathrm{e} \frac{z^2}{\beta^2} \left[ F(\beta) - \delta(\beta) \right],
\end{equation}
the main term for ionizing electrons is given by
\begin{equation} 
F(\gamma) = \ln \frac{m_\mathrm{e}^2 c^4 (\gamma^3-\gamma^2-\gamma+1)}{2 I^2} + \frac{8 \ln 2 + 9}{8 \gamma^2} -\frac{8 \ln 2 +1}{4 \gamma} +\frac{1}{8}
\end{equation}
with the relations $\gamma = 1/\sqrt{1-\beta^2}$, $E = m_\mathrm{e} c^2 (\gamma -1)$, $m_\mathrm{e} c^2 = \qty{0.511}{MeV}$, and $I = \qty{78}{eV}$ for water \citep{ICRU_2016}.
The density effect has been modeled to
\begin{equation}
\delta(X) = \begin{cases}
0 & (X \leq X_0) \\
2 \ln(10)\,X + a\,(X_1-X)^m + C & (X_0 < X \leq X_1) \\
2 \ln(10)\,X + C & (X > X_1) 
\end{cases}
\end{equation}
as a function of $X = \lg(\beta\gamma)$ with the parameters $X_0 = 0.2400$, $X_1 = 2.8004$, $a = 0.09116$, $m = 3.4773$, and $C = -3.5017$ for water \citep{Sternheimer_1984}.
Figure \ref{fig_C1} shows the electronic stopping power $S_\mathrm{e}$ and the relative density effect $\delta/F$ at relevant electron energies.

\begin{figure}[htb]
\includegraphics[width=\columnwidth]{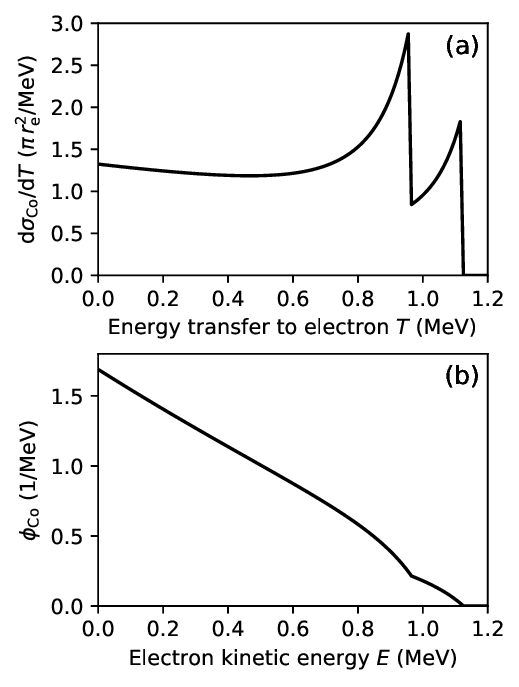}
\caption{(a) Compton scattering cross section derivative of energy transfer to electron for primary \gamma-ray photons per $^{60}$Co decay, $\mathrm{d}\sigma_\mathrm{Co}/\mathrm{d}T$, and (b) energy spectrum of these electrons in matter at ideal equilibrium, $\phi_\mathrm{Co}(E)$.}
\label{fig_C2}
\end{figure}

The Klein--Nishina formula \citep{Klein_1929} gives the Compton scattering cross section for a photon of energy $E_\text{\gamma}$ to transfer energy $T$ to an electron at rest,
\begin{equation}
\frac{\mathrm{d}\sigma}{\mathrm{d}T} = \frac{\text{\pi} r_\mathrm{e}^2 m_\mathrm{e} c^2}{E_\text{\gamma}^2} \left[
2 - \frac{T}{E_\text{\gamma}} \frac{2 m_\mathrm{e} c^2-T}{E_\text{\gamma}-T} 
+ \left( \frac{T}{E_\text{\gamma}} \frac{m_\mathrm{e} c^2}{E_\text{\gamma}-T} \right)^2 
\right].
\end{equation}
The electron production cross section for primary \gamma-ray photons per $^{60}$Co decay is then
\begin{equation}
\frac{\mathrm{d}\sigma_\mathrm{Co}}{\mathrm{d}T} = \left.\frac{\mathrm{d}\sigma}{\mathrm{d}T}\right|_{E_\text{\gamma} =\qty{1.17}{MeV}} + \left.\frac{\mathrm{d}\sigma}{\mathrm{d}T}\right|_{E_\text{\gamma} =\qty{1.33}{MeV}}.
\end{equation}
The produced electrons slow down and stop in matter.
At ideal equilibrium, where the number of electrons balances between production and termination, the energy spectrum or probability density as a function of kinetic energy $E$ is given by
\begin{equation}
\left. \phi_\mathrm{Co}(E) \propto \int_E^\infty \frac{\mathrm{d}\sigma_\mathrm{Co}}{\mathrm{d}T} \mathrm{d}T \quad \middle | \quad \int_0^\infty \phi_\mathrm{Co}(E)\, \mathrm{d}E = 1. \right.
\end{equation}
Figure \ref{fig_C2} shows the electron production cross section $\mathrm{d}\sigma_\mathrm{Co}/\mathrm{d}T$ and the electron energy spectrum in matter, $\phi_\mathrm{Co}(E)$.

The relative density effect in the dosimetry of $^{60}$Co \gamma\ rays in water at electron equilibrium was evaluated by dose-weighted averaging,
\begin{equation}
\left<\frac{\delta}{F}\right>_\mathrm{Co} = \left. \int_0^\infty \frac{\delta}{F}\,S_\mathrm{e}\, \phi_\mathrm{Co}\, \mathrm{d}E \middle/ \int_0^\infty S_\mathrm{e}\, \phi_\mathrm{Co}\, \mathrm{d}E \right. \approx \qty{0.13}{\%},
\end{equation}
which is small enough to be disregarded in dosimetry.
Furthermore, in reality, photon energy reduction by prior Compton scattering and electron non-equilibrium due to photon attenuation in matter will shift the electron spectrum in Fig.~\ref{fig_C2} toward lower energies. 
This spectrum shift will suppress the density effect at high energies depicted in Fig.~\ref{fig_C1}.
We therefore conclude that WEG for proton and ion beams will also be valid for $^{60}$Co \gamma\ rays.
This justifies the cancellation of ${{s_\mathrm{e}}_\mathrm{w/weg}}_{Q/Q_0}$ in Eq.~(\ref{eq_Deweg}).
\bibliographystyle{elsarticle-harv}
\bibliography{references}

\end{document}